\documentclass[12pt]{article}
\usepackage{graphicx}
\usepackage[cp1251]{inputenc}
\usepackage{rotating}
\begin{document}
 \noindent {\footnotesize\it Astronomy Reports, 2020, Vol. 64, No 4, pp. 326--335}
 \newcommand{\dif}{\textrm{d}}

 \noindent
 \begin{tabular}{llllllllllllllllllllllllllllllllllllllllllllll}
 & & & & & & & & & & & & & & & & & & & & & & & & & & & & & & & & & & & & & \\\hline\hline
 \end{tabular}

 \vskip 0.5cm
  \centerline{\bf\large New Expansion Rate Estimate of the Scorpius-Centaurus Association}
  \centerline{\bf\large Based on T Tauri Stars from the Gaia DR2 Catalog}
 \bigskip
 \bigskip
  \centerline
 {
 V.V. Bobylev and A.T. Bajkova
 }
 \bigskip
 \centerline{\small \it
 Central (Pulkovo) Astronomical Observatory, Russian Academy of Sciences,}
 \centerline{\small \it Pulkovskoe shosse 65, St. Petersburg, 196140 Russia}
 \bigskip
 \bigskip
 \bigskip

 {
{\bf Abstract}---The kinematic properties of the
Scorpius–Centaurus association were studied using spatial
velocities of approximately 700 young T Tauri stars. Their proper
motions and trigonometric parallaxes were selected by Zari et al.
from the Gaia DR2 catalog, and radial velocities were taken from
various sources. The linear expansion coefficient’s new estimate
of the association $K=39\pm2$~km/s/kpc is obtained by considering
the influence of the galactic spiral density wave with an
amplitude of radial disturbances $f_R=5$~km/s and solar phase in
the wave $-120^\circ$. The proper rotation of the association is
shown to be small. The residual velocity ellipsoid of these stars
has semimajor axes
$\sigma_{1,2,3}=(7.72,1.87,1.74)\pm(0.56,0.37,0.22)$~km/s and is
positioned at an angle $12\pm2^\circ$ to the galactic plane.
  }

\medskip DOI: 10.1134/S1063772920040022

 \section{INTRODUCTION}
The Scorpius–Centaurus stellar association (Sco OB2) is a typical,
not very young OB association. It is located near the Sun and has
been studied in more detail than other similar structures. For
example, according to the data of the modern Gaia DR2 catalog [1,
2], the Scorpius–Centaurus association includes approximately 3000
main sequence candidate members and more than 11 000 T Tauri stars
[3]. The association is divided into three groups: Upper Scorpius
(US), Upper Centaurus–Lupus (UCL), and Lower Centaurus–Crux (LCC)
with mean distances of 145, 140, and 118 pc, respectively [4, 5].

Compared to open clusters, stellar associations have a
significantly lower density. Ambartsumian [6] hypothesized their
gravitational instability and gradual dissipation. Blaauw was
among the first who modeled [7] and estimated [8] the expansion
effect of a stellar association. In particular, using the data on
young massive stars of spectral class B, he found the linear
expansion coefficient of the Centaurus–Crux association
$K=50$~km/s/kpc, which allowed the expansion time to be estimated
at 20 Myr. This kinematic estimate is in rather good agreement
with modern age estimates of the main association members: US
(below 10 Ma), UCL (16--20 Ma) and LCC (16--20 Ma), which were
obtained by fitting the evolutionary tracks of stars to isochrones
on the Hertzsprung–Russell diagram and other methods [9--11].

Analysis of modern kinematic data made it possible to detect the
effects of proper rotation and expansion in other known OB
associations [12, 13]. For example, the Per OB1 and Car OB1
associations expand at a rate of approximately 6 km/s [14, 15].

Using the example of the Scorpius–Centaurus association, Sartori
et al. [10] showed that there were no differences in the
distribution and kinematics between massive and low-mass (T Tauri)
stars of comparable age. They proposed a model for the formation
of this association as a result of the action of a galactic spiral
density wave on the gas–dust proto-cloud. In [16], a detailed
model of sequential star formation was developed as applied to the
Scorpius–Centaurus association.

Bobylev and Bajkova [18] obtained an estimate of the linear
expansion coefficient of the Scorpius–Centaurus association
$K=46\pm8$~km/s/kpc based on a sample of young massive stars from
the HIPPARCOS catalog [17]. Torres et al. [19] showed that a
significant number of young star groups from a wide neighborhood
of the Scorpius–Centaurus association are affected by the
expansion effect at approximately the same rate. In the analysis
of the OB star kinematics, the authors of [20] suggested that the
galactic spiral density wave can have a significant effect on the
determination of the $K$-effect in the Gould belt and in the
Scorpius–Centaurus association.

This study’s aim is to determine the spatial and kinematic
characteristics of a young star system belonging to the
Scorpius–Centaurus association, to refine the known expansion
effect of this association, to estimate the proper system
rotation, and to analyze the residual velocities of the stars
calculated with allowance for the galactic spiral density wave.

 \section{DATA}
Zari et al. [21] created a compilation catalog of T Tauri stars.
These stars were selected from the Gaia DR2 catalog according to
kinematic and photometric characteristics. All the stars are
located no further than 500 pc from the Sun, since the sample’s
radius was limited to $\pi<2$~milliarcseconds (mas). The vast
majority of the stars belong to the Gould belt. The stars were
selected by their proper motions by analyzing the smoothed point
distribution on the plane $\mu_\alpha\cos\delta\times\mu_\delta$
using the restriction on the tangential velocity of a star
$\sqrt{(\mu_\alpha\cos\delta)^2+\mu^2_\delta}<40$~km/s. The radial
velocities in the catalog [21] were taken from various sources, in
particular, from the Gaia DR2 catalog. However, the number of the
stars with radial velocities is substantially smaller than that of
the stars with proper motions.

In this study, we use T Tauri stars from the sample most closely
related to the Gould belt. It is denoted pmsvt3 in the catalog
[21] and contains 23686 stars with proper motions and parallaxes,
as well as approximately 2000 stars with radial velocities.

Stars with relative parallax errors of less than 15\% belonging to
the Scorpius–Centaurus association were selected from the pmsvt3
sample. The new sample includes about 5300 candidate members with
proper motions and parallaxes. Radial velocities are measured for
approximately 700 of these stars. During the selection, the stars
with latitudes $b$ from $-15^\circ$ to $35^\circ$ were taken in
accordance with the association map (e.g., [5]). In addition, the
following restrictions were used: heliocentric distance
$r<220$~pc, coordinate is in the range from $-50$ to 250~pc, and
coordinate from $-200$ to 50~pc.

 \section{METHODS}
A rectangular coordinate system centered at the Sun was used,
where the $x$ axis is oriented toward the galactic center, the $y$
axis toward the galactic rotation, and the $z$ axis toward the
north galactic pole. Then, $x=r\cos l\cos b,$ $y=r\sin l\cos b$
and $z=r\sin b.$

We know three components of a star's velocity from the
observations: radial velocity $V_r$ and two tangential velocity
projections $V_l=4.74r\mu_l\cos b$ and $V_b=4.74r\mu_b,$ which are
oriented along the galactic longitude $l$ and latitude $b$,
respectively, and expressed in km/s. Here, the coefficient 4.74 is
the ratio of the kilometer number in an astronomical unit to the
number of seconds in a tropical year, and $r=1/\pi$ is the
heliocentric distance of the star in kpc, which is calculated
through the star's parallax in mas. The proper motion components
and are expressed in mas/year.

Spatial velocities $V_r, V_l$ and $V_b$ are calculated through the
components $U, V$ and $W,$; velocity $U$ is oriented from the Sun
to the Galaxy’s center, $V$ in the direction of the Galaxy’s
rotation, and $W$ to the north galactic pole:
\begin{equation}
 \begin{array}{lll}
 U=V_r\cos l\cos b-V_l\sin l-V_b\cos l\sin b,\\
 V=V_r\sin l\cos b+V_l\cos l-V_b\sin l\sin b,\\
 W=V_r\sin b                +V_b\cos b.
 \label{UVW}
 \end{array}
 \end{equation}
Obviously, to calculate spatial velocities, it is necessary to use
stars with complete information, i.e., with known parallax, radial
velocity, and proper motion components.

Figure 1 shows the selected star distribution with known radial
velocities projected onto the galactic plane $xy$ and their
spatial velocities. The figure also shows 33 stars with velocities
$V>10$~km/s, which were not used in solving the basic kinematic
equations. As can be seen from the figure, the selected stars form
a very compact cluster in the $UV$ plane of velocities (Fig. 1b)
and are of great interest for kinematic analysis.

 \begin{figure} {\begin{center}
 \includegraphics[width=160mm]{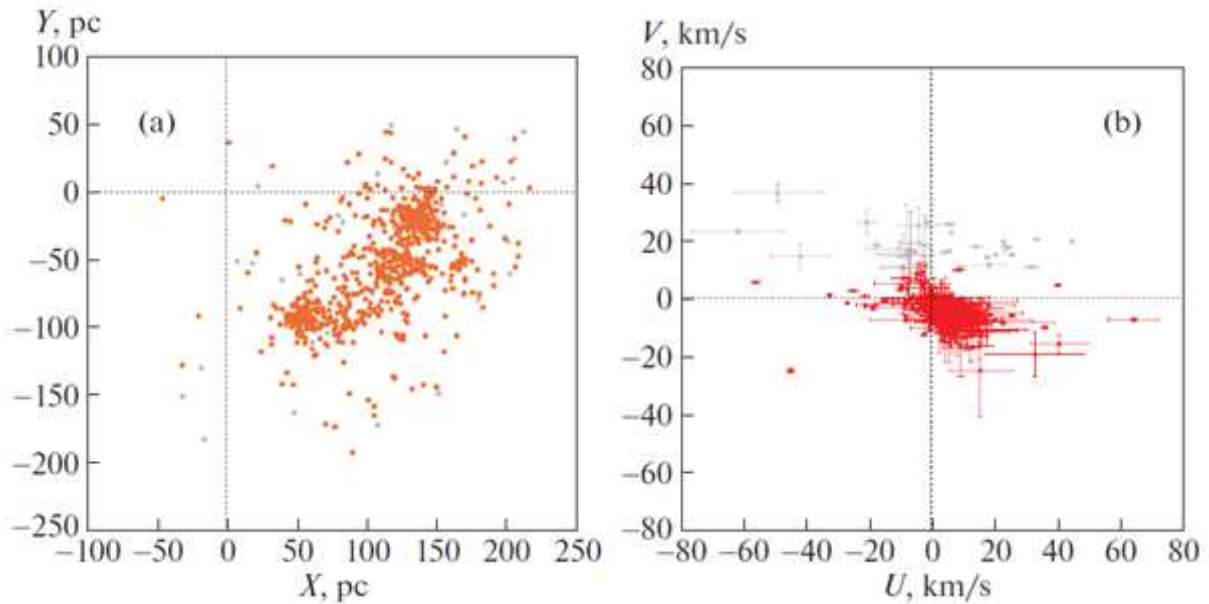}
 \caption{
(a) Distribution of the probable members of the Scorpius–Centaurus
association on the galactic plane $xy$; (b) spatial $UV$
velocities of these stars. The velocities are given relative to
the LSR; the bright dots indicate the stars with velocities
$V>10$~km/s.
  }
 \label{fXY}
 \end{center} } \end{figure}

 \subsection{Residual Velocities}
In the residual velocity formation, the Sun’s peculiar velocity
relative to the local standard of rest (LSR), $U_\odot,$ $V_\odot$
and $W_\odot$ is primarily considered. The radial extension of the
region under study (along $R$) is less than 200 pc; thus, there is
no particular need to consider the Galaxy’s differential rotation,
especially since the galactic rotation’s curve is nearly flat.
However, it is interesting to consider the influence of the
galactic spiral density wave. The expressions for taking these two
effects into account are as follows:
 \begin{equation}
 \begin{array}{lll}
 V_r=V^*_r-[-U_\odot\cos b\cos l-V_\odot\cos b\sin l-W_\odot\sin b\\
 +\tilde{v}_\theta \sin(l+\theta)\cos b-\tilde{v}_R\cos(l+\theta)\cos b],
 \label{EQU-1}
 \end{array}
 \end{equation}
 \begin{equation}
 \begin{array}{lll}
 V_l=V^*_l-[U_\odot\sin l-V_\odot\cos l-r\Omega_0\cos b
 + \tilde{v}_\theta \cos(l+\theta)
 + \tilde{v}_R\sin(l+\theta) ],
 \label{EQU-2}
 \end{array}
 \end{equation}
  \begin{equation}
 \begin{array}{lll}
 V_b=V^*_b-[U_\odot\cos l\sin b + V_\odot\sin l \sin b
 -W_\odot\cos b\\
 - \tilde{v}_\theta \sin(l+\theta)\sin b
 + \tilde{v}_R \cos(l+\theta)\sin b ],
 \label{EQU-3}
 \end{array}
 \end{equation}
where $V^*_r,V^*_l,V^*_b$ in the right-hand parts of the equations
are the initial uncorrected velocities, and $V_r,V_l,V_b$ in the
left-hand parts are the corrected velocities, using the ones in
which the residual velocities $U,V,W$ based on relations (1) can
be calculated; $R$ is the distance from the star to the galaxy’s
rotation axis, $R^2=r^2\cos^2 b-2R_0 r\cos b\cos l+R^2_0.$

The distance $R_0$ is taken as $8.0\pm0.15$~kpc [22]. The specific
values of the Sun’s peculiar velocity relative to the LSR,
$(U_\odot,V_\odot,W_\odot)=(11.1,12.2,7.3)$~km/s, are taken in
accordance with the definition [23].

There are two other velocities that are also of interest: the
radial velocity $V_R$ oriented from the galactic center along the
radius and the velocity $V_\theta$ orthogonal to it and directed
along the galaxy’s rotation. These velocities are calculated based
on the following relationships:
 \begin{equation}
 \begin{array}{lll}
  V_\theta=~~U\sin\theta+(V_0+V)\cos\theta, \\
       V_R= -U\cos\theta+(V_0+V)\sin\theta,
 \label{VRVT}
 \end{array}
 \end{equation}
where the position angle $\theta$ satisfies the relation
$\tan\theta=y/(R_0-x)$, $x,y,z$ are the rectangular heliocentric
coordinates of the star; and $V_0$ is the linear velocity of the
Galaxy’s rotation at a solar distance $R_0$.

Considering the influence of the spiral density wave, we used a
model based on the linear theory of density waves by Lin and Shu
[24], in which the potential’s perturbation has the form of a
traveling wave.
  \begin{equation}
 \begin{array}{lll}
 \tilde{v}_R = f_R \cos \chi,\\
 \tilde{v}_\theta = f_\theta \sin \chi,\\
 \chi= m[\cot (i)\ln (R/R_0)-\theta]+\chi_\odot,
 \label{wave}
 \end{array}
 \end{equation}
where $f_R$ and $f_\theta$ are the perturbation amplitudes of the
radial and azimuthal velocities; $i$ is the twist angle of spirals
($i<0$ for twisting spirals); $m$ is the number of arms;
$\chi_\odot$ is the phase angle of the Sun, which is measured from
the center of the Carina–Sagittarius Arm in this study; $\lambda$
is the distance (along the galactocentric radial direction)
between the adjacent segments of spiral arms in a near-solar
neighborhood (spiral wave length), which is calculated from the
relation
 \begin{equation}
 \tan (i)=\lambda m/(2\pi R_0).
 \label{tani}
 \end{equation}
The described method for considering the influence of a spiral
density wave was used, for example, in [25] or [26].

There is currently no certainty regarding the number of spiral
arms $m$ in our galaxy. However, in this case, the neighborhood
under consideration is small, and the position angle
$\theta\rightarrow0^\circ$ in formula (6). Under this condition,
it can be easily shown that the inclusion of a spiral wave does
not depend on $m$. Indeed, in accordance with relation (7), we
have $\cot (i)=2\pi R_0/\lambda m.$. After substituting into (6),
$m$ is canceled. Thus, considering the influence of a spiral
density wave, it is necessary to have four parameters: $\lambda$,
$f_R$, $f_\theta$ and $\chi_\odot.$

The assessment efficiency depends both on the perturbation
amplitude of the spiral wave and on the solar phase in the wave.
For example, at a certain phase value, a zero effect can be
obtained even with a large amplitude. On the other hand, for a
certain amplitude of radial perturbations, the value of the linear
expansion coefficient $K$ from the stellar system decreases by the
same amount. This was obtained in the kinematics analysis of the
Gould belt [27].

In this paper, in accordance with the analysis of various star
samples [28--31], the following parameters of the spiral density
wave are adopted:
 $\lambda=2.2$~kpc, $f_R=5$~km/s,
 $f_\theta=0$~km/s and $\chi_\odot=-120^\circ.$
Since very young stars are considered, studies focused on the
parameter values of the spiral density wave that were obtained by
various authors for the youngest objects. For example, the value
of the Sun’s phase in the density wave was found to be
$\chi_\odot=-125\pm10^\circ$ from the analysis of maser sources
with measured trigonometric parallaxes [29],
$\chi_\odot=-120\pm10^\circ$ from a sample of young open star
clusters [30], and $\chi_\odot=-121\pm3^\circ$ from Cepheids [28].
The amplitudes of radial disturbance velocities $f_R$, as a rule,
significantly differ from zero, while those of tangential
$f_\theta$ do not [29, 30]. A current definition summary of $f_R$
and $f_\theta$ is provided by Loktin and Popova (Table 2 in [31]).
Based on open-star cluster data from the current version of the
``Uniform Catalog of Parameters of Open Star Clusters'' and data
from the Gaia DR2 catalog, these authors found
$f_R=4.6\pm0.7$~km/s and $f_\theta=1.1\pm0.4$~km/s [31]. The value
of the disturbance wavelength in the Sun’s vicinity lies in the
range of 2.0--2.5 kpc [29--31]; it is well determined by both the
positions of the stars and their kinematics. For example, the
values $\lambda_\theta=2.3\pm0.5$~kpc and
$\lambda_R=2.2\pm0.5$~kpc were found from the analysis of both
tangential and radial velocities of young open star clusters from
the Gaia DR2 catalog [30].

 \subsection{Residual Velocities Ellipsoid}
To determine the parameters of the residual velocity ellipsoid of
the stars, the following known method [32] is used. In the classic
version, six moments of the second order $a,b,c,f,e,$ and $d$ are
considered:
\begin{equation}
 \begin{array}{lll}
 a=\langle U^2\rangle-\langle U^2_\odot\rangle,\qquad\quad
 b=\langle V^2\rangle-\langle V^2_\odot\rangle,\qquad\quad
 c=\langle W^2\rangle-\langle W^2_\odot\rangle,\\
 f=\langle VW\rangle-\langle V_\odot W_\odot\rangle,\quad
 e=\langle WU\rangle-\langle W_\odot U_\odot\rangle,\quad
 d=\langle UV\rangle-\langle U_\odot V_\odot\rangle,
 \label{moments}
 \end{array}
 \end{equation}
As noted above, if necessary, the observable velocities can be
released not only from the Sun’s peculiar motion, but also from
the Galaxy’s differential rotation or from the influence of the
spiral density wave. The moments $a,b,c,f,e,$ and $d$ are the
coefficients of the surface equation
 \begin{equation}
 ax^2+by^2+cz^2+2fyz+2ezx+2dxy=1,
 \end{equation}
as well as the symmetric tensor components of residual velocity
moments
 \begin{equation}
 \left(\matrix {
  a& d & e\cr
  d& b & f\cr
  e& f & c\cr }\right).
 \label{ff-5}
 \end{equation}
The values of all elements of this tensor can be determined from
the solution of the following system of conditional equations:
\begin{equation}
 \begin{array}{lll}
 V^2_l= a\sin^2 l+b\cos^2 l\sin^2 l -2d\sin l\cos l,
 \label{EQsigm-2}
 \end{array}
 \end{equation}
\begin{equation}
 \begin{array}{lll}
 V^2_b= a\sin^2 b\cos^2 l+b\sin^2 b\sin^2 l +c\cos^2 b\\
 -2f\cos b\sin b\sin l -2e\cos b\sin b\cos l
 +2d\sin l\cos l\sin^2 b,
 \label{EQsigm-3}
 \end{array}
 \end{equation}
\begin{equation}
 \begin{array}{lll}
 V_lV_b= a\sin l\cos l\sin b +b\sin l\cos l\sin b\\
 +f\cos l\cos b-e\sin l\cos b +d(\sin^2 l\sin b-\cos^2\sin b),
 \label{EQsigm-4}
 \end{array}
 \end{equation}
\begin{equation}
 \begin{array}{lll}
 V_b V_r=-a\cos^2 l\cos b\sin b -b\sin^2 l\sin b\cos b+c\sin b\cos b\\
 +f(\cos^2 b\sin l-\sin l\sin^2 b)
 +e(\cos^2 b\cos l-\cos l\sin^2 b)\\
 -d(\cos l\sin l\sin b\cos b +\sin l\cos l\cos b\sin b),
 \label{EQsigm-5}
 \end{array}
 \end{equation}
\begin{equation}
 \begin{array}{lll}
 V_l V_r=-a\cos b\cos l\sin l+b\cos b\cos l\sin l\\
    +f\sin b\cos l-e\sin b\sin l
    +d(\cos b\cos^2 l-\cos b\sin^2 l).
 \label{EQsigm-6}
 \end{array}
 \end{equation}
The solution is sought using the least squares method (LSM) with
respect to six unknowns $a,b,c,f,e,$ and $d$. The eigenvalues of
tensor (10) $\lambda_{1,2,3}$ are then found from the solution of
the secular equation
 \begin{equation}
 \left|\matrix
 {
a-\lambda&          d&        e\cr
       d & b-\lambda &        f\cr
       e &          f&c-\lambda\cr
 }
 \right|=0.
 \label{ff-7}
 \end{equation}
The eigenvalues of this equation are equal to the inverse values
of the ellipsoid’s squared semiaxes of the velocity moments and,
at the same time, to the ellipsoid’s squared semiaxes of the
residual velocities:
 \begin{equation}
 \begin{array}{lll}
 \lambda_1=\sigma^2_1, \lambda_2=\sigma^2_2, \lambda_3=\sigma^2_3,\\
 \lambda_1>\lambda_2>\lambda_3.
 \end{array}
 \end{equation}
Directions of the principal axes of tensor (16) $L_{1,2,3}$ and
$B_{1,2,3}$ are found from the relations
 \begin{equation}
 \tan L_{1,2,3}={{ef-(c-\lambda)d}\over {(b-\lambda)(c-\lambda)-f^2}},
 \label{ff-41}
 \end{equation}
 \begin{equation}
 \tan B_{1,2,3}={{(b-\lambda)e-df}\over{f^2-(b-\lambda)(c-\lambda)}}\cos L_{1,2,3}.
 \label{ff-42}
 \end{equation}
The errors of determining and are estimated according to the
following procedure:
 \begin{equation}
 \renewcommand{\arraystretch}{2.8}
  \begin{array}{lll}
  \displaystyle
 \varepsilon (L_2)= \varepsilon (L_3)= {{\varepsilon (\overline
 {UV})}\over{a-b}},\quad
  \displaystyle
 \varepsilon (B_2)= \varepsilon (\varphi)={{\varepsilon (\overline
 {UW})}\over{a-c}},\quad
  \displaystyle
 \varepsilon (B_3)= \varepsilon (\psi)= {{\varepsilon (\overline {VW})}\over{b-c}},\\
  \displaystyle
 \varepsilon^2 (L_1)={\varphi^2 \varepsilon^2 (\psi)+\psi^2 \varepsilon^2
 (\varphi)\over{(\varphi^2+\psi^2)^2}},\quad
  \displaystyle
 \varepsilon^2 (B_1)= {\sin^2 L_1 \varepsilon^2 (\psi)+\cos^2 L_1 \varepsilon^2 (L_1)\over{(\sin^2 L_1+\psi^2)^2}},
 \label{ff-65}
  \end{array}
 \end{equation}
where $ \varphi=\cot B_1 \cos L_1$ and $\psi=\cot B_1 \sin L_1.$
In this case, it is necessary to calculate beforehand three values
 $\overline {U^2V^2}$, $\overline {U^2W^2},$ and $\overline {V^2W^2},$ then
 \begin{equation}
 \renewcommand{\arraystretch}{1.6}
  \begin{array}{lll}
  \displaystyle
 \varepsilon^2 (\overline {UV})= (\overline{U^2V^2}-d^2)/n, \quad
  \displaystyle
 \varepsilon^2 (\overline {UW})= (\overline {U^2W^2}-e^2)/n, \quad
  \displaystyle
 \varepsilon^2 (\overline {VW})= (\overline {V^2W^2}-f^2)/n,
 \label{ff-73}
  \end{array}
 \end{equation}
where $n$ is the number of stars. Here, the errors of each axis
are estimated independently, with the exception of $L_2$ and
$L_3,$ in which its errors are calculated by one formula.

 \subsection{Kinematic Model}
From the analysis of residual velocities $V_r,V_l,$ and $V_b$ the
average group velocity $(U,V,W)_\diamondsuit,$ can be determined,
as well as four analogues of Oort constants
$(A,B,C,K)_\diamondsuit.$ In this case, these characterize the
effects of proper rotation ($A_\diamondsuit$ and
$B_\diamondsuit$), as well as expansion and compression
($C_\diamondsuit$ and $K_\diamondsuit$), of the low-mass star
sample. A simple kinematic model similar to the Oort–Lindblad
model is used in this case [32]:
 \begin{equation}
 \begin{array}{lll}
 V_r= U_\diamondsuit\cos b\cos l+
      V_\diamondsuit\cos b\sin l+
      W_\diamondsuit\sin b\\
      +rA_\diamondsuit\cos^2 b\sin 2l
      +rC_\diamondsuit\cos^2 b \cos 2l
      +rK_\diamondsuit\cos^2 b,
 \label{MOD-1}
 \end{array}
 \end{equation}
 \begin{equation}
 \begin{array}{lll}
 V_l=
    -U_\diamondsuit\sin l
    +V_\diamondsuit\cos l
 +rA_\diamondsuit\cos b\cos 2l-rC_\diamondsuit\cos b\sin 2l
 +rB_\diamondsuit\cos b,
 \label{MOD-2}
 \end{array}
 \end{equation}
 \begin{equation}
 \begin{array}{lll}
 V_b=
    -U_\diamondsuit\cos l\sin b
    -V_\diamondsuit\sin l\sin b
   + W_\diamondsuit\cos b\\
 -rA_\diamondsuit\sin b\cos b\sin 2l
 -rC_\diamondsuit\cos b\sin b\cos 2l
 -rK_\diamondsuit\cos b\sin b.
 \label{MOD-3}
 \end{array}
 \end{equation}
The unknowns $(U,V,W)_\diamondsuit$ and $(A,B,C,K)_\diamondsuit$
result from a joint LSM solution of the conditional equation
system (22)--(24). The following weight system is used:
$w_r=S_0/\sqrt {S_0^2+\sigma^2_{V_r}},$
 $w_l=S_0/\sqrt {S_0^2+\sigma^2_{V_l}}$ and
 $w_b=S_0/\sqrt {S_0^2+\sigma^2_{V_b}},$
where $S_0$ is the ``space'' dispersion; and $\sigma_{V_r},
\sigma_{V_l},$ and $\sigma_{V_b}$ are the dispersions of the
errors of the corresponding observed velocities. The value of
$S_0$ is comparable to the mean-square residual (unit weight
error) when solving conditional equations of the form (22)--(24).
For the analysis of star residual velocities in this study, $S_0$
is taken as 3 km/s. The 3$\sigma$ criterion is applied for
discarding residuals.

Using the parameter values $A_\diamondsuit$ and $C_\diamondsuit,$
the $l_{xy}$ angle (vertex deviation) in accordance with the
relation proposed by Parenago [33] is calculated:
  \begin{equation}
  \tan (2l_{xy})= {{(AK-BC)_\diamondsuit} \over {(AB+KC)_\diamondsuit}},
  \label{lxyK}
  \end{equation}
in which the absence of expansion or compression (at $K=0$) takes
a more familiar (as in the analysis of galactic rotation) form
$\tan (2l_{xy})=-C_\diamondsuit/A_\diamondsuit.$ In the case of
pure rotation, angle $l_{xy}$ points exactly at the kinematic
center.

Note several important relationships in our kinematic model [32]:
  \begin{equation}
   \begin{array}{lll}
    (\Omega_0)_\diamondsuit=(B-A)_\diamondsuit,\\
        (V'_0)_\diamondsuit=(B+A)_\diamondsuit,
   \label{Omega}
  \end{array}
 \end{equation}
where $(\Omega_0)_\diamondsuit$ is the rotation’s angular
velocity, and $(V'_0)_\diamondsuit=(\partial V_\theta/\partial
R)_\diamondsuit$ is the first derivative of the linear rotational
velocity $(V_\theta)_\diamondsuit$ at the point $R=R_0.$

For the angular expansion and compression rate
$(k_0)_\diamondsuit$ and the first derivative of the linear radial
velocity (oriented along the radius from the system’s kinematic
center) of expansion and compression $V_R$ at the point $R=R_0$,
studies show [32]
  \begin{equation}
   \begin{array}{lll}
       (k_0)_\diamondsuit=(K-C)_\diamondsuit,\\
    (V'_R)_\diamondsuit=(K+C)_\diamondsuit.
   \label{kkk}
  \end{array}
 \end{equation}

 \section*{RESULTS AND DISCUSSION}
Table 1 lists the results obtained by a joint LSM solution of the
form’s conditional equations (22)--(24). The first column shows
the sought-for parameters and their associated values; the second
column gives the results obtained from the stars corrected for the
Sun’s peculiar motion relative to the LSR; the third column gives
the results obtained from the same stars with additional
corrections to proper motions and radial velocities for the
influence of a spiral density wave with amplitude $f_R=5$~km/s and
the solar phase in the wave $\chi_\odot=-120^\circ$; the fourth
column gives the results obtained from the stars described in the
previous step, but all longitudes are corrected by $l=l-l_{xy},$
i.e., reduced to the new kinematic center of the system. In this
new coordinate system, the values of the parameters
$(U,V,W)_\diamondsuit$ and $(v,l,b)_\diamondsuit$ are not of great
interest and thus are not listed here.

The lower part of the table shows the parameters of the residual
velocity ellipsoid found as a LSM solution result of the
conditional equation system (11)--(15). Based on the analysis of
the Gaia catalog data by Wright and Mamajek (2018), the following
average values of velocity dispersions were determined for the
three main groups in the Scorpius–Centaurus association:
$3.20^{+0.22}_{-0.20}$~km/s (US),
 $2.45^{+0.20}_{-0.20}$~km/s (UCL) and
 $2.15^{+0.47}_{-0.24}$~km/s (LCC). As observed in Table 1, the unit
weight errors $\sigma_0$ (it has the sense of the average error
over the three coordinate axes), as well as the dispersion values
of the residual velocity ellipsoids, are in good agreement with
the indicated estimates.

The second column in Table 1 gives the group velocity vector
components of the sample relative to the Sun $({\overline
U},{\overline V},{\overline
W})=(-10.7,-16.1,-6.2)\pm(0.4,0.4,0.1)$~km/s. These were
calculated as a simple average of the star velocities that were
not corrected in any way.

In [18], the value $({\overline U},{\overline V},{\overline
W})=(-11.8,-18.2,-6.1)\pm(0.8,0.8,0.3)$~km/s was obtained from the
analysis of 134 bright members of the Scorpius–Centaurus
association from the HIPPARCOS catalog [17]. Goldman et al. [35]
determined $({\overline U},{\overline V},{\overline
W})=(-8.2,-20.9,-6.1)\pm(1.1,1.5,0.6)$~km/s for 487 stars of
subgroup C from LCC with an average age of 10 Ma. Wright and
Mamajek [5] found $({\overline U},{\overline V},{\overline
W})=(-7.2,-19.6,-6.1)\pm(0.2,0.2,0.1)$~km/s based on the analysis
of approximately 250 bright members of the Scorpius–Centaurus
association from the Gaia DR2 catalog.

Due to the fact that 33 stars with velocities $V>10$~km/s were
discarded (Fig. 1), the sample is kinematically very homogeneous.
Earlier, an estimate of the linear expansion coefficient
$K=46\pm8$~km/s/kpc was obtained from a sample of approximately
200 massive OB stars from the HIPPARCOS catalog [18]. Now, we
obtain a significantly smaller error in determining this
coefficient, $\pm$2 km/s/kpc. The value $K=41\pm2$~km/s/kpc from
the second column of Table 1 is in close agreement with the
analysis of massive stars that was also obtained in [18] without
considering the spiral density wave.

The value $l_{xy}=-13\pm12^\circ$ (third column of Table 1)
indicates that the kinematic center of the star system lies in the
fourth galactic quadrant. However, the error in determining this
quantity is large. On the other hand, the first axis direction of
the residual velocity ellipsoid of the stars $L_1=324\pm5^\circ$
(lower part of Table 1) is determined with a smaller error. Both
$l_{xy}$ and $L_1$ alignments are counted from the direction
toward the galactic center $l=0^\circ$ but in opposite ways. For
their correct comparison, we will count from the direction
$l=0^\circ$ uniformly. For example, for $L_1=324\pm5^\circ$, we
can write $324^\circ-360^\circ=-36^\circ$. Thus, the direction
$-36\pm5^\circ$ can be considered the direction toward the
kinematic center of the Scorpius–Centaurus association.

 \begin{table}[t]
 \caption[]{\small
Parameters of the Oort–Lindblad kinematic model (upper part) and
parameters of the residual velocity ellipsoid (lower part) }
  \begin{center}  \label{t:01}   \small
  \begin{tabular}{|c|r|r|r|r|r|r|r|}\hline
  Parameters   &  Before accounting  & After accounting &  $l_{new}=l-l_{xy}$ \\\hline

   ${\overline U},$ km/s & $-10.67\pm0.40$ & & $$ \\
   ${\overline V},$ km/s & $-16.14\pm0.39$ & & $$ \\
   ${\overline W},$ km/s & $ -6.19\pm0.08$ & & $$ \\

  $N_\star$        &          697 &         697  &        697  \\
  $\sigma_0,$ km/s &          2.2 &         2.2  &        2.2  \\
   $U_\diamondsuit,$ km/s   & $ 0.43\pm0.40$ & $ 2.55\pm0.41$ & \\
   $V_\diamondsuit,$ km/s   & $-3.94\pm0.39$ & $-3.43\pm0.40$ & \\
   $W_\diamondsuit,$ km/s   & $ 1.11\pm0.08$ & $ 1.11\pm0.08$ & \\
    $v_\diamondsuit,$ km/s  & $4.12\pm0.38$  & $ 4.42\pm0.39$ & \\
    $l_\diamondsuit,$ deg & $276\pm6$      & $  307\pm5$    & \\
    $b_\diamondsuit,$ deg & $ 16\pm2$      & $   15\pm2$    & \\
  $A_\diamondsuit,$ km/s/kpc & $ 2.5\pm2.0$ & $ 2.0\pm2.0$ & $-0.1\pm2.0$ \\
  $B_\diamondsuit,$ km/s/kpc & $-2.0\pm1.8$ & $ 0.8\pm1.9$ & $ 0.8\pm1.9$ \\
  $C_\diamondsuit,$ km/s/kpc & $ 3.9\pm1.9$ & $-4.3\pm1.9$ & $-4.8\pm2.0$ \\
  $K_\diamondsuit,$ km/s/kpc & $41.1\pm2.1$ & $36.8\pm2.1$ & $36.8\pm2.1$ \\
        $l_{xy},$ deg        & $ 18\pm12$   &  $-13\pm12$  &       $0$    \\
  \hline
  $\sigma_1,$ km/s  & $7.71\pm0.62$ & $8.23\pm0.58$ & \\
  $\sigma_2,$ km/s  & $2.40\pm0.38$ & $2.18\pm0.41$ & \\
  $\sigma_3,$ km/s  & $1.86\pm0.24$ & $1.58\pm0.28$ & \\
    $L_1, B_1,$ deg & $323\pm7,~ ~3\pm1$ & $324\pm5,~ ~2\pm1$ & \\
    $L_3, B_3,$ deg & $ 56\pm9,~ 39\pm3$ & $ 60\pm7,~ 68\pm3$ & \\
    $L_2, B_2,$ deg & $230\pm9,~ 51\pm2$ & $233\pm7,~ 22\pm2$ & \\
  \hline
  \end{tabular}\end{center} \end{table}
 \begin{figure}[t] {\begin{center}
 \includegraphics[width=140mm]{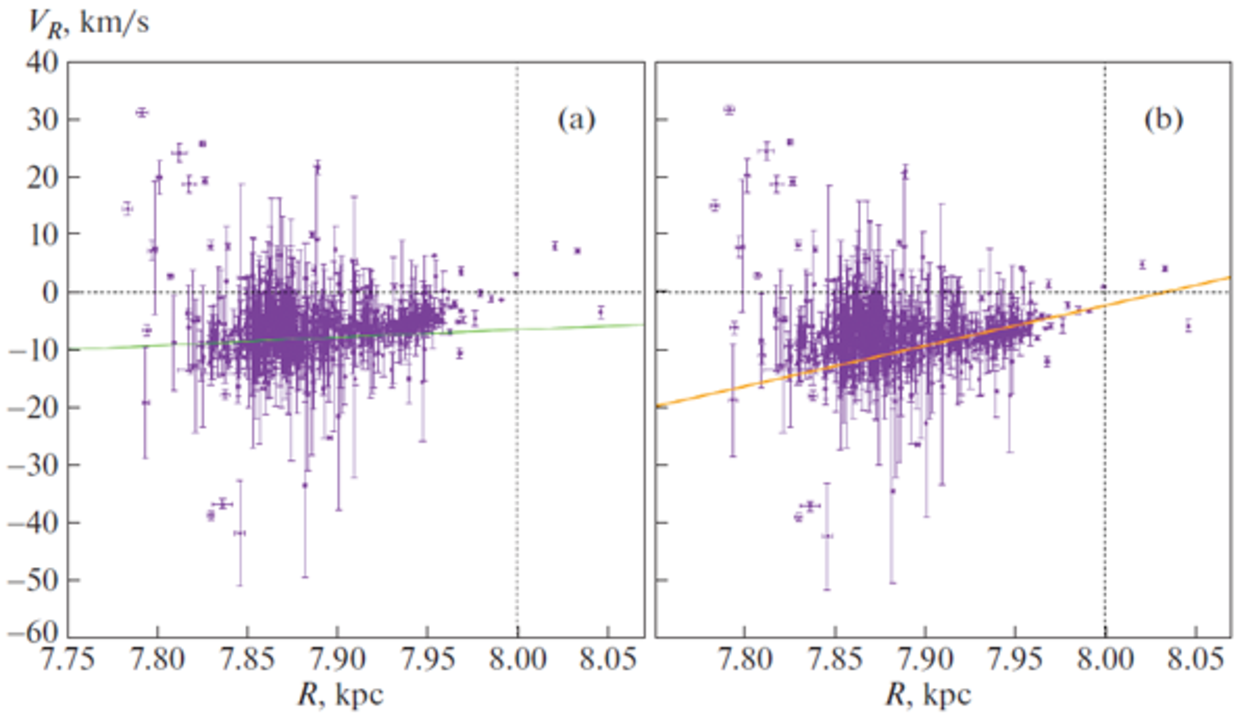}
 \caption{
Galactocentric radial velocities of stars $V_R$ depending on the
distance $R$ (a) corrected for the Sun’s motion relative to the
LSR (the green line shows the influence of the spiral density
wave) and (b) corrected for the Sun’s motion relative to the LSR
with the influence of the spiral density wave taken into account
(the orange line corresponds to $K$ effect found from these
stars). }
 \label{fRad}
 \end{center} } \end{figure}

In the coordinate system with a new center (fourth column of Table
1), the constants $A_\diamondsuit$ and $C_\diamondsuit$ do not
differ significantly from zero.

Figure 2 illustrates the stellar velocities $V_R$ depending on
distance $R$. The velocities are corrected for the Sun’s motion
relative to the LSR. Figure 2a shows a wave
 $$
  -5\cos\biggl[-{2\pi R_0\over 2.2}\ln\biggl({R\over R_0}\biggr)-120^\circ\biggr],
 $$
written down in accordance with relations (6) and (7), with the
perturbation amplitude $f_R=5$~km/s, wavelength $\lambda=2.2$~kpc,
and the solar phase in the wave $\chi_\odot=-120^\circ$; the minus
sign before the formula means that the perturbation is oriented
toward the center of the Galaxy in the center of a spiral arm (for
example, when $R\approx7.2$~kpc).

By definition, $2K=V_R/R+\partial V_R/\partial R$ if the
rotational velocity $V_\theta$ is independent of the angle
$\theta,$ $\partial V_\theta/\partial\theta=0$~[32]. At a constant
angular velocity, i.e., at $\partial V_R/\partial R=0$, $\partial
V_R/\partial R=0$ and $2K=V_R/R$. Figure 2b shows the dependence
$V_R/R=2K$ with the value $K=37$~km/s/kpc.

Based on the solution from the fourth column of Table 1 using
relation (27),  $K-C=41.6\pm2.8$~km/s/kpc and
$K+C=32.0\pm2.8$~km/s/kpc are found. From an observer’s
perspective, these values show that there is a large angular
velocity of expansion and a large positive derivative of the
expansion’s linear velocity.

It is interesting to determine the kinematic parameters for US,
UCL, and LCC separately. To do so, the entire sample was divided
into three thirty-degree sectors by galactic longitude $l$. The
results of solving equations (22)--(24), by allowing a spiral wave
to influence three samples, are given in Table 2.

As can be seen in Table 2, the value of the linear expansion
coefficient $K=13.6\pm6.6$~km/s/kpc determined from the star
samples with longitudes $l<300^\circ$ is very different from other
results. These stars also have the highest unit weight error
$\sigma_0.$ Thus, there is a strong influence from the stars of
the first galactic quadrant, which are most likely background
stars. Therefore, further analysis was carried out without these
stars.

As a result, the LSM solution of conditional equations (22)--(24)
was obtained under the condition $l>300^\circ$ using 574 stars,
where the mean motion vector is
$(U,V,W)_\diamondsuit=(2.96,-2.84,1.24)\pm(0.45,0.44,0.08)$~km/s
(this vector corresponds to the total velocity
$v_\diamondsuit=4.28\pm0.43$~km/s with direction
$l_\diamondsuit=316\pm6^\circ$ and $b_\diamondsuit=17\pm3^\circ$),
as well as the following values of the remaining parameters:
 \begin{equation}
 \begin{array}{lll}
  A_\diamondsuit=-0.1\pm2.1~\hbox{km/s/kpc}, \\
  B_\diamondsuit= 1.5\pm2.0~\hbox{km/s/kpc}, \\
  C_\diamondsuit=-9.5\pm2.1~\hbox{km/s/kpc}, \\
  K_\diamondsuit=39.1\pm2.3~\hbox{km/s/kpc}, \\
    l_{xy}=-1\pm6^\circ,
 \label{best solution-1}
 \end{array}
 \end{equation}
where the unit weight error $\sigma_0$ amounted to 1.9 km/s. Based
on the obtained value of the linear expansion coefficient
$K=39\pm2$~km/s/kpc, the characteristic time of expansion of the
complex can be estimated according to the well-known formula
$T=977.5/K$, $T=25\pm2$ which is Myr.

To determine the rotation parameters, the star radial velocities
are not necessary; two equations (23)--(24) or even one (23) are
sufficient. Thus, to study the proper rotation of the
Scorpius–Centaurus association, it would be better to use a
catalog with a huge number of stars from Damiani et al. [3] when
this directory appears in the Strasbourg database.

 \begin{table}[t]
 \caption[]{\small
Parameters of the Oort–Lindblad kinematic model found from the
samples in three longitude intervals }
  \begin{center}  \label{t:02}   \small
  \begin{tabular}{|c|r|r|r|r|r|r|r|}\hline
  Parameters       & $l<300^\circ$ &  $l:300-330^\circ$ &  $l>330$ \\\hline

  $N_\star$        &          116 &         213  &        367  \\
  $\sigma_0,$ km/s &          3.5 &         1.4  &        2.1  \\
   $U_\diamondsuit,$ km/s   & $ 4.5\pm1.1$ & $ 3.9\pm0.6$ & $ 4.0\pm0.7$ \\
   $V_\diamondsuit,$ km/s   & $-3.1\pm1.1$ & $-5.1\pm0.6$ & $-1.8\pm0.6$ \\
   $W_\diamondsuit,$ km/s   & $ 0.4\pm0.3$ & $ 1.3\pm0.1$ & $ 1.2\pm0.1$ \\
    $v_\diamondsuit,$ km/s  & $5.5\pm1.1$  & $ 6.5\pm0.6$ & $ 4.6\pm0.6$ \\
    $l_\diamondsuit,$ deg & $326\pm12$   & $  307\pm5$  & $  335\pm8$ \\
    $b_\diamondsuit,$ deg & $  4\pm4$    & $   12\pm2$  & $   15\pm5$ \\
  $A_\diamondsuit,$ km/s/kpc & $  3.1\pm6.1$ & $ 15.2\pm3.6$ & $-11.7\pm4.1$ \\
  $B_\diamondsuit,$ km/s/kpc & $-11.3\pm5.9$ & $-11.7\pm3.6$ & $ 10.2\pm4.1$ \\
  $C_\diamondsuit,$ km/s/kpc & $-20.9\pm6.3$ & $ 12.1\pm3.8$ & $-25.9\pm3.9$ \\
  $K_\diamondsuit,$ km/s/kpc & $ 13.6\pm6.6$ & $ 37.6\pm3.9$ & $ 41.4\pm4.0$ \\
   $l_{xy},$ deg         & $ 16\pm8$     &    $34\pm5$   & $    5\pm4$   \\
  \hline
  \end{tabular}\end{center} \end{table}

As noted in the Data section, the sample contains approximately
5300 candidate members of the Scorpius– Centaurus association with
their proper motions and parallaxes. The radial velocities are
known for a small star fraction in this sample. In this case, the
LSM solution of the three equation system (22)--(24) is followed
as: a star with proper motions gives two equations (23) and (24);
if the radial velocity is available, the star gives all three
equations. With this approach, the rotation parameters are the
main focus, assuming that the expansion parameter is already
determined reliably.

As a result, the following solution was obtained:
$(U,V,W)_\diamondsuit=(3.11,-1.88,1.10)\pm(0.17,0.13,0.02)$~km/s
($v_\diamondsuit=3.80\pm0.15$~km/s with direction
 $l_\diamondsuit=329\pm2^\circ$ and $b_\diamondsuit=17\pm1^\circ$),
and
 \begin{equation}
 \begin{array}{lll}
  A_\diamondsuit= 1.8\pm0.8~\hbox{km/s/kpc}, \\
  B_\diamondsuit=-7.5\pm0.6~\hbox{km/s/kpc}, \\
  C_\diamondsuit=-9.6\pm0.6~\hbox{km/s/kpc}, \\
  K_\diamondsuit=33.1\pm0.9~\hbox{km/s/kpc}, \\
          l_{xy}=   1\pm2^\circ,
 \label{best solution-2}
 \end{array}
 \end{equation}
where $\sigma_0$ amounted to 1.3 km/s. Unlike solution (28), here,
the values of all four constants are determined more accurately.
The value $l_{xy}$ is close to zero, so there is no need to use a
new coordinate system.

From the solution (29) with the dominant expansion, the following
rotation parameters using the relation (26) are found:
$(\Omega_0)_\diamondsuit=-9.3\pm1.0$~km/s/kpc and
$(V'_0)_\diamondsuit=-5.7\pm1.0$~km/s/kpc. The sign of this
angular velocity indicates that its direction coincides with
galactic rotation. At the point of the observer, the modulus of
this velocity increases.

For the expansion, using relation (27), we find the following
parameters:
 $(k_0)_\diamondsuit=42.7\pm1.1$~km/s/kpc and
 $(V'_R)_\diamondsuit=24.5\pm1.1$~km/s/kpc. Thus, from the observer’s
perspective, the modulus of this velocity also increases.

Applying a similar approach to solution (28), as well as to the
solution in the last column of Table 1, similar values for the
expansion parameters and the absence of proper rotation are found.
Thus, the expansion parameters are determined reliably, and the
parameters of proper rotation strongly depend on the adopted
restrictions. However, in general, the proper rotation of the
Scorpius–Centaurus association is small.

Fern\'andez et al. [34] traced the kinematic evolution of the
Scorpius–Centaurus association by analyzing the galactic orbits of
various association parts in the past. The orbits were constructed
in an axisymmetric potential with additional allowance for the
spiral density wave. The expansion of the association was
confirmed. Concurrently, the Local Bubble evolution was studied as
well.

The studies by Wright and Mamajek [5] are also important to note.
They tested the kinematics of the Scorpius–Centaurus association
using several methods. In particular, they considered a method for
searching for the linear expansion coefficient by the radial
velocities of stars, and also reconstructed the star orbits in
order to find the time point of their spatial concentration’s
smallest area. The authors concluded that there was no evidence
that the subgroups under consideration had a more compact
configuration in the past. In other words, they found no expansion
signs of the association. Conversely, Goldman and et al. [35]
showed the presence of an expansion of the star subsystem in the
association and in LCC with a linear expansion coefficient
$K\sim35$~km/s/kpc.

The following parameters of the residual velocity ellipsoid were
found from the stars with radial velocities that were used to
search for solution (28):
 \begin{equation}
 \begin{array}{lll}
  \sigma_1= 7.72\pm0.56~\hbox{km/s}, \\
  \sigma_2= 1.87\pm0.37~\hbox{km/s}, \\
  \sigma_3= 1.74\pm0.22~\hbox{km/s}
 \label{rezult-6}
 \end{array}
 \end{equation}
and orientation parameters of this ellipsoid
 \begin{equation}
  \matrix {
  L_1=323\pm7^\circ, & B_1=~~3\pm1^\circ, \cr
  L_2=~54\pm6^\circ, & B_2=~12\pm2^\circ, \cr
  L_3=305\pm6^\circ, & B_3=~78\pm2^\circ. \cr
   }
 \label{rezult-66}
 \end{equation}
The direction $L_1=323^\circ (-37^\circ)$ first agrees well with
the direction towards the geometric center of the
Scorpius-Centaurus association (Fig. 1) and, secondly, agrees well
with the value $l_{xy}=-44^\circ$ found on the basis of the
Oort–Lindblad model (solution (28)). In contrast to the ellipsoids
of Table 1, it is interesting to note that this ellipsoid’s
orientation is in excellent agreement with the Gould belt’s
orientation [36]. For example, on the basis of a similar approach
in Bobylev’s study [27], it was shown that the residual velocity
ellipsoid of the Gould belt stars has principal semiaxes
$\sigma_{1,2,3}=(8.9,5.6,3.0)\pm(0.1,0.2,0.1)$~km/s and it is
located at an angle $22\pm1^\circ$ to the galactic plane.

 \section{CONCLUSIONS}
The kinematics of young T Tauri stars belonging to the
Scorpius–Centaurus association was studied. For this purpose, the
catalog of this star type was used, with their proper motions and
parallaxes by Zari et al. [21] based on the Gaia DR2 catalog. The
catalog [21] contains more than 23 000 stars and is dedicated to
the stars belonging to the Gould belt.

Present studies mainly focus on the search for stars closely
related to the Scorpius–Centaurus association. For this purpose,
restrictions were placed on both the coordinates of the stars and
their spatial velocities. The Oort–Lindblad kinematic model was
used as the main model, in which the sample’s vector of the
average group velocity relative to the LSR $(U,V,W)_\diamondsuit$
was determined, as well as four analogues of Oort constants
$(A,B,C,K)_\diamondsuit$.

An important feature found in this study is the influence
assessment of the spiral density wave. The assessment efficiency
depends both on the amplitude of the perturbation of the spiral
wave and on the solar phase in the wave. For example, at a certain
value of the phase, a zero effect can be obtained even for a large
amplitude. On the other hand, for a certain amplitude of radial
perturbations, the linear expansion’s determined coefficient value
of the stellar system $K_\diamondsuit$ decreases by the same
amount. We have chosen a relatively small value of the radial
perturbation amplitude and a zero value of the tangential
perturbation amplitude found earlier from masers, OB stars, and
young Cepheids.

First, a kinematic analysis was performed for about 700
Scorpius–Centaurus stars for which the proper motions, parallaxes,
and radial velocities were measured. The new estimate of the
linear expansion coefficient $K_\diamondsuit=39\pm2$~km/s/kpc was
obtained considering the Sun’s peculiar velocity relative to the
LSR and the influence of the galactic spiral density wave with the
amplitude of radial perturbations $f_R=5$~km/s and the solar phase
in the wave $-120^\circ.$ Compared to the expansion, the proper
rotation of the Scorpius–Centaurus association is small; it is
poorly determined since the rotation parameters strongly depend on
the adopted restrictions.

The residual velocity ellipsoid of the stars from this sample has
the principal semiaxes
$\sigma_{1,2,3}=(7.72,1.87,1.74)\pm(0.56,0.37,0.22)$~km/s and is
positioned at an angle of $12\pm2^\circ$ to the galactic plane.
The orientation of the ellipsoid shows that it lies close to the
Gould plane, and its first axis lies in the direction
$143^\circ-323^\circ$.

Using approximately 5300 stars with their proper motions and
parallaxes, it was shown that, relative to the LSR, the
Scorpius–Centaurus association moves at a velocity
$v_\diamondsuit=3.80\pm0.15$~km/s in the direction
$l_\diamondsuit=329\pm2^\circ$ and $b_\diamondsuit=17\pm1^\circ.$

 \subsubsection*{ACKNOWLEDGEMENTS}
The authors would like to thank the reviewer for useful comments,
which helped improve the article.

 \subsubsection*{FUNDING}
The study was supported in part by the Program of the Presidium of
the Russian Academy of Sciences KP19--270 ``Questions of origin
and evolution of the Universe using the methods of ground-based
observations and space research.''

 \medskip\subsubsection*{REFERENCES}

 {\small
  \quad ~1. A.G.A. Brown, A. Vallenari, T. Prusti,  et al. (Gaia Collab.),
Astron. Astrophys. 616, 1 (2018).

2. L. Lindegren, J. Hernandez, A. Bombrun, et al. (Gaia Collab.),
Astron. Astrophys. 616, 2 (2018).

3. F. Damiani, L. Prisinzano, I. Pillitteri, et al., Astron.
Astrophys. 623, 112 (2019).

4. P.T. de Zeeuw, R. Hoogerwerf, J.H.J. de Bruijne, et al.,
Astron. J. 117, 354 (1999).

5. N.J. Wright and E.E. Mamajek, Mon. Not. R. Astron. Soc. 476,
381 (2018).

6. V.A. Ambartsumian, Astron. Zh. 26 (3) (1949).

7. A. Blaauw, Bull. Astron. Inst. Netherland 11, 414 (1952).

8. A. Blaauw, Ann. Rev. Astron. Astrophys. 2, 213 (1964).

9. E. E. Mamajek, M. Meyer, and J. Liebert, Astron. J. 124, 1670
(2002).

10. M.J. Sartori, J.R.D. Lepine, and W. S. Dias, Astron.
Astrophys. 404, 913 (2003).

11. T. Preibisch and E. Mamajek, in Handbook of Star Forming
Regions, Ed. by Bo Reipurth, Vol. 5 of The Southern Sky ASP
Monograph Publications (ASP, San Francisco, 2008), Vol. 2.

12. T. Cantat-Gaudin, C. Jordi, N.J. Wright,  et al., Astron.
Astrophys. 626, 17 (2019).

13. A. Rao, P. Gandhi, C. Knigge, et al., arXiv: 1908.00810
(2019).

14. A.M. Mel’nik and A.K. Dambis, Mon. Not. R. Astron. Soc. 472,
3887 (2017).

15. A.M. Mel’nik and A.K. Dambis, Astron. Rep. 62, 998 (2018).

16. T. Preibisch and H. Zinnecker, Astron. J. 117, 2381 (1999).

17. The HIPPARCOS and Tycho Catalogues, ESA SP– 1200 (1997).

18. V.V. Bobylev and A.T. Bajkova, Astron. Lett. 33, 571 (2007).

19. C.A.O. Torres, R. Quast, C.H.F. Melo, and M.F. Sterzik, in
Handbook of Star Forming Regions, Ed. by Bo Reipurth, Vol. 5 of
The Southern Sky ASP Monograph Publications (ASP, San Francisco,
2008), Vol. 2.

20. V.V. Bobylev and A.T. Bajkova, Astron. Lett. 39, 532 (2013).

21. E. Zari, H. Hashemi, A.G.A. Brown, et al., Astron. Astrophys.
620, 172 (2018).

22. T. Camarillo, M. Varun, M. Tyler, et al., Publ. Astron. Soc.
Pacif. 130, 4101 (2018).

23. R. Sch\"onrich, J. Binney, and W. Dehnen, Mon. Not. R. Astron.
Soc. 403, 1829 (2010).

24. C.C. Lin and F.H. Shu, Astrophys. J. 140, 646 (1964).

25. Yu. N. Mishurov and I.A. Zenina, Astron. Astrophys. 341, 81
(1999).

26. D. Fern\'andez, F. Figueras, and J. Torra, Astron. Astrophys.
372, 833 (2001).

27. V.V. Bobylev, Astron. Lett. 46 (2020, in press).

28. A.K. Dambis, L.N. Berdnikov, Yu. N. Efremov, et al., Astron.
Lett. 41, 489 (2015).

29. A.S. Rastorguev, M.V. Zabolotskikh, A.K. Dambis, et al.,
Astrophys. Bull. 72, 122 (2017).

30. V.V. Bobylev and A.T. Bajkova, Astron. Lett. 45, 151 (2019).

31. A. V. Loktin and M. E. Popova, Astrophys. Bull. 74, 270
(2019).

32. K.F. Ogorodnikov, Dynamics of Stellar Systems (Pergamon,
Oxford, 1965).

33. P.P. Parenago, Course of Stellar Astronomy (Gosizdat, Moscow,
1954) [in Russian].

34. D. Fern\'andez, F. Figueras, and J. Torra, Astron. Astrophys.
480, 735 (2008).

35. B. Goldman, S. Roser, E. Schilbach, et al., Astrophys. J. 868,
32 (2018).

36. V.V. Bobylev, Astrophysics 57, 583 (2014).

 }
 \end{document}